\documentclass[conference]{IEEEtran}
\usepackage{psfig}      
\usepackage{algorithmic}
\usepackage{algorithm}
\usepackage{graphicx}
\usepackage{tabularx}
\usepackage{amsmath}
\usepackage{color}
\usepackage{epsfig}
\usepackage{amssymb}
\usepackage{subfigure}

\newcommand{\Define}{\stackrel{\triangle}{=}}
\newcommand{\mc}{\mathcal}
\newcommand{\mbf}{\mathbf}
\newcommand{\beq}{\begin{equation}}
\newcommand{\eeq}{\end{equation}}
\newcommand{\beqa}{\begin{eqnarray}}
\newcommand{\eeqa}{\end{eqnarray}}
\newcommand{\hsp}{\hspace}

\newcommand{\nn}{\nonumber}

\begin{document}
\twocolumn

\title{\LARGE Achievable Rate Region of Quantized Broadcast and MAC Channels}
\author{Suresh Chandrasekaran$^{\dagger}$, Saif K. Mohammed$^{\ddagger}$, 
and A. Chockalingam$^{\dagger}$\\
{\normalsize $^{\dagger}$Department of ECE, Indian Institute of Science,
Bangalore 560012, India}\\
{\normalsize $^{\ddagger}$Communication Systems Division, Department of 
Electrical Engg., Link\"oping University, Sweden}
\vspace{-5mm}
}
\maketitle
\thispagestyle{empty}

\begin{abstract}
In this paper, we study the achievable rate region of Gaussian multiuser 
channels with the messages transmitted being from finite input alphabets 
and the outputs being {\em quantized at the receiver}. In particular, we 
focus on the achievable rate region of $i)$ Gaussian broadcast channel 
(GBC) and $ii)$ Gaussian multiple access channel (GMAC).
First, we study the achievable rate region of two-user GBC when the messages 
to be transmitted to both the users take values from finite signal sets and 
the received signal is quantized at both the users. We refer to this 
channel as {\em quantized broadcast channel (QBC)}. We observe that the 
capacity region defined for a GBC does not carry over as such to QBC. We 
show that the optimal decoding scheme for GBC (i.e., high SNR user doing
successive decoding and low SNR user decoding its message alone) is not
optimal for QBC. We then propose an achievable rate region for QBC based
on two different schemes. We present achievable rate region results for 
the case of uniform quantization at the receivers.
Next, we investigate the achievable rate region of two-user GMAC with  
finite input alphabet and quantized receiver output. We refer to this 
channel as {\em quantized multiple access channel (QMAC)}. We derive 
expressions for the achievable rate region of a two-user QMAC. We show 
that, with finite input alphabet, the achievable rate region with the 
commonly used uniform receiver quantizer has a significant loss compared 
to the achievable rate region without receiver quantization. We propose a 
{\em non-uniform quantizer} which has a significantly larger rate region 
compared to what is achieved with a uniform quantizer in QMAC.
\end{abstract}

{\em {\bfseries Keywords} -- {\footnotesize Gaussian broadcast channel,
Gaussian multiple access channel, finite input alphabet, quantized 
receiver, achievable rate region, successive decoding, discrete memoryless 
channel.}}

\section{Introduction}
\label{sec_intro}
Communication receivers are often based on digital signal processing, where
the analog received signal is quantized into finite number of bits using
analog-to-digital converters (ADC) whose outputs are then processed in
digital domain. These ADCs are expected to operate at high speeds in order
to meet the increasing throughput and bandwidth requirements. However, at
high conversion speeds, the precision of ADCs is typically low which results
in loss of system performance \cite{walden99}. For example, low-precision
receiver quantization can cause floors in the bit error performance 
\cite{ashu},\cite{mezg}. Also, it has been shown that in a
single-input single-output (SISO) point-to-point single user system with
additive white Gaussian noise (AWGN), low-precision receiver quantization
results in significant loss of capacity when compared to an unquantized
receiver \cite{singh09}. Motivated by the increasing need to investigate
the effect of receiver quantization in high-throughput communication, we,
in this paper, address the issue of characterizing the achievable rate
region of two different Gaussian multiuser channels, namely,
\begin{enumerate}
\item 	Gaussian broadcast channel (GBC) with finite input alphabet and
	quantized receiver output; we refer to this channel as
	the {\em Quantized broadcast channel (QBC)}, 
\item 	Gaussian multiple access channel (GMAC) with finite input alphabet 
	and quantized receiver output; we refer to this channel as
	{\em Quantized multiple access channel (QMAC)}, 
\end{enumerate}
\vspace{-2mm}
and report some interesting results.

GBC comes under the class of degraded
broadcast channels, for which capacity is known. For a two-user GBC,
it is known that the capacity is achieved when superposition coding
is done at the transmitter assuming that the users' messages are from
Gaussian distribution, and, at the receiver, the high SNR user does
successive decoding and the low SNR user decodes its message alone
considering the other user's message as noise \cite{gamal}. However,
the capacity region of two-user QBC is not known. Recently, achievable 
rate region for two-user GBC when the input messages are from finite 
signal sets and the received signals are {\em unquantized} has been 
studied in \cite{naveen}, and it is referred to as the constellation 
constrained (CC) capacity of GBC \cite{biglieri}.

Our present contribution first gives achievable rate region for two-user 
QBC in Section \ref{sec_qbc}. The main results on QBC are summarized as 
follows.
\begin{itemize}
\item	The capacity region defined for a GBC does not carry over as such
	to QBC.
\item	Once quantization is done at the receiver in a GBC, the channel
	is no more degraded. Therefore, the optimal decoding scheme
	for GBC (i.e., high SNR user alone doing successive decoding)
	does not necessarily result in achievable rate pairs for QBC.
\item	We then propose achievable rate region for QBC based on two 
	different  schemes (scheme 1 and scheme 2). In scheme 1, user 1 
	will do successive decoding and user 2 will not, whereas, in 
	scheme 2, user 2 will do successive decoding and user 1 will not. 
	In addition to this, in both the schemes, the message for the user 
	which does not do successive decoding is coded at such a rate that 
	the message of that user can be decoded error free at both the 
	receivers.
\item	Rotation of one of the user's input alphabet with respect to
	the other user's alphabet marginally enlarges the achievable
	rate region of QBC when almost equal powers are allotted to
	both the users.
\end{itemize}

Next, in Section \ref{sec_qmac}, we address the achievable rate region 
of two-user QMAC. With finite input alphabets and an {\em unquantized} 
receiver, the two-user GMAC rate region has been studied in \cite{harsh08}. 
In \cite{harsh08}, in terms of the achievable rate region, it was shown 
that, compared to having both the users transmit using the same finite 
signal set, it is better to have the second user transmit using a rotated 
version of the first user's signal set. We refer to the two-user GMAC 
system model in \cite{harsh08} (with finite input alphabet and no output 
quantization) as constellation constrained MAC (CCMAC). 

In this paper, instead of assuming an unquantized receiver as 
was done in \cite{harsh08}, we consider quantized receiver. Since uniform 
quantizers are commonly used in communication receivers, we first consider 
uniform quantization at the receiver, and show that with uniform quantization, 
there is a significant reduction in the achievable rate region compared to 
the CCMAC rate region. This is due to the fact that the received analog 
signal is densely distributed around the origin, and is therefore not 
efficiently quantized with a uniform quantizer. This then motivates us to 
propose a {\em non-uniform quantizer} with finely spaced quantization 
intervals near the origin. We show that the proposed non-uniform quantizer 
results in enlargement of the achievable rate region of two-user QMAC 
compared to that achieved with a uniform quantizer. It is further observed 
that, with increasing number of users, the probability distribution of the 
received analog signal is more and more dense around the origin. Hence, it 
is expected that with increasing number of users, larger enlargement in 
rate region of QMAC may be achieved with non-uniform quantization compared 
to uniform quantization. 

The rest of this paper is organized as follows. Achievable rate region 
of two-user QBC is studied in Section \ref{sec_qbc}. Achievable rate region 
of two-user QMAC is presented in Section \ref{sec_qmac}. Conclusions 
are given in Section \ref{sec_concl}.

\section{Quantized Broadcast Channel}
\label{sec_qbc}
In this section, we propose achievable rate region for two-user QBC. We 
show achievable rate region results when the users employ uniform 
receiver quantization.

\subsection{System Model }
\label{sec_sys_mod}
\begin{figure}
\begin{center}
\epsfysize=7.25cm
\epsfxsize=7.00cm
\epsfbox{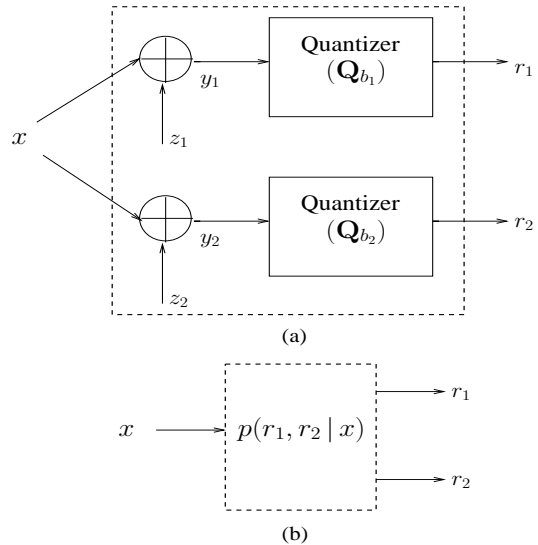}
\vspace{-1mm}
\caption{(a) Two-user Gaussian broadcast channel with receiver quantization. 
(b) Equivalent discrete memoryless channel.}
\label{fig_2user}
\end{center}
\vspace{-4mm}
\end{figure}

We consider a two-user GBC as shown in Fig. \ref{fig_2user}. Let $x_1$ 
and $x_2$ denote the messages to be transmitted to the users 1 and 2, 
respectively. Let $x_1$ and $x_2$ take values from finite signal sets 
$\mc{X}_1$ and $\mc{X}_2$, respectively. The sets $\mc{X}_1$ and 
$\mc{X}_2$ contain $N_1$ and $N_2$ equi-probable complex entries, 
respectively. Let the sum signal set of $\mc{X}_1$ and $\mc{X}_2$ be 
defined as
\begin{eqnarray}
\mc{X}=\{x_1+x_2 \mid x_1 \in \mc{X}_1, x_2 \in \mc{X}_2\}. 
\label{eq_sumset}
\end{eqnarray}
Let $X^I$ and $X^Q$ be defined as
\begin{equation}
X^I \,\, \Define \,\, \max_{a \in \mc{X}} |a^I|, \quad
X^Q \,\, \Define \,\, \max_{a \in \mc{X}} |a^Q|,
\label{eq_xiq}
\end{equation}
where $a^I$ and $a^Q$ represent the real and imaginary components of $a$, 
respectively.

Let $x\in\mc{X}$ be the message sent by the transmitter to the users 1 
and 2 with an average power constraint $P$. We further assume that the 
average power constraint on $x_1$ is $\alpha P$ and the average power 
constraint on $x_2$ is $(1-\alpha)P$, where $\alpha \in (0,1)$. Let 
$z_1 \sim \mc{CN}(0, \sigma_1^2)$ and  $z_2 \sim \mc{CN}(0, \sigma_2^2)$  
denote the additive white Gaussian noise at receivers 1 and 2, 
respectively. The signal-to-noise ratio (SNR) at user 1 
(SNR1) is $P/\sigma_1^2$ and the SNR at user 2 (SNR2) is $P/\sigma_2^2$.
The received signal at user 1 is then given by
\begin{eqnarray}
y_1 \,\, = \,\,  x+z_1 \,\, = \,\, x_1+x_2+z_1.
\end{eqnarray}
Similarly, the received signal at user 2 is given by
\begin{eqnarray}
y_2 \,\,= \,\, x+z_2 \,\,= \,\, x_1+x_2+z_2.
\end{eqnarray}
The received analog signals, $y_1$ at user 1 and $y_2$ at user 2, are 
quantized independently, resulting in outputs $r_1$ at user 1 and $r_2$ 
at user 2. The complex quantizer at each user is composed of two similar 
quantizers acting independently on the real and imaginary components of 
the received analog signal. The real and imaginary components of the 
quantized output for the users 1 and 2 are then given by
\begin{eqnarray}
r_1^{I}\,\, = \,\ \mathbf{Q}_{b_1}(y_1^I), \quad r_1^{Q} \,\,= \,\,\mathbf{Q}_{b_1}(y_1^Q),
\label{eq_r1rirq}\\
r_2^{I} \,\,= \,\,\mathbf{Q}_{b_2}(y_2^I), \quad r_2^{Q} \,\,= \,\,\mathbf{Q}_{b_2}(y_2^Q),
\label{eq_r2rirq}
\end{eqnarray}
where the functions $\mbf{Q}_{b_1}(.)$ and $\mbf{Q}_{b_2}(.)$ model the 
quantizers having a resolution of $b_1$ and $b_2$ bits, respectively. The 
function $\mbf{Q}_{b_1}(.)$ defines a mapping from the set of real numbers 
$\mathbb{R}$ to a finite alphabet set $\mc{S}_{b_1}$ of cardinality
$2^{b_1}$, i.e., 
\begin{equation}
\mbf{Q}_{b_1}: \mathbb{R} \mapsto \mc{S}_{b_1}, \quad \mc{S}_{b_1} \subset \mathbb{R},
\quad  |\mc{S}_{b_1}|=2^{b_1}.
\label{eq_Qb1}
\end{equation}
Similarly,
\begin{equation}
\mbf{Q}_{b_2}: \mathbb{R} \mapsto \mc{S}_{b_2}, \quad \mc{S}_{b_2} \subset \mathbb{R},
\quad  |\mc{S}_{b_2}|=2^{b_2}.
\label{eq_Qb2}
\end{equation}
Thus, the quantized received signals $r_1$ at user 1 and $r_2$ at user 2 
take values from the sets $\mc{R}_1$ and $\mc{R}_2$, respectively, where
\begin{eqnarray}
&\mc{R}_1 = \{r_1^I+jr_1^Q \;| \; r_1^I,r_1^Q\in\mc{S}_{b_1} \}, & |\mc{R}_1|=2^{2b_1},\\
&\mc{R}_2 = \{r_2^I+jr_2^Q \;| \; r_2^I,r_2^Q\in\mc{S}_{b_2} \}. & |\mc{R}_2|=2^{2b_2}.
\end{eqnarray}
Henceforth, we refer to the above system model as {\em quantized broadcast 
channel (QBC)}.

\subsection{Achievable Rate Region of QBC} 
\label{sec_rate}
In this subsection, we derive analytical expressions for the achievable 
rate region of two-user QBC. 

The capacity region of a two-user GBC is known \cite{cover99},\cite{cover_bc}, 
and is given by the set of all rate pairs $(R_1,R_2)$ satisfying
\begin{eqnarray}
R_1 & \leq & I(x_1;y_1 \, | \, x_2) \label{eq_gr1}\\
R_2 & \leq & I(x_2;y_2), \label{eq_gr2}
\end{eqnarray}
assuming $\sigma_1^2 < \sigma_2^2$, where $R_1$ and $R_2$ represent the 
rates achieved by user 1 and user 2, respectively. The optimal input 
distribution that attains the capacity 
is known to be {\em Gaussian}. The optimal decoding scheme is that, 
user 1 does successive decoding (i.e., user 1 first decodes user 2's 
message assuming its own message as noise and subtracts the decoded 
user 2's message $\hat{x}_2$ from its received signal $y_1$, and then 
decodes its own message from the subtracted signal $y_1-\hat{x}_2$), and 
user 2 decodes its message alone by considering user 1's message as noise. 
This GBC belongs to a class of broadcast channel, {\em degraded broadcast 
channel}, which satisfies the condition
\begin{eqnarray}
p(y_1,y_2|x)& = & p(y_1|x) \; p(y_2|y_1), 
\label{eq_deg}
\end{eqnarray}
i.e., $x\rightarrow y_1 \rightarrow y_2$ (Markov).
However, observe that, in QBC,
\begin{eqnarray}
p(r_1,r_2|x) & \neq & p(r_1|x) \; p(r_2|r_1), 
\label{eq_degn}
\end{eqnarray}
i.e., $x\rightarrow r_1 \rightarrow r_2$ is not true. Hence, the effective 
channel $(\mc{X},p(r_1,r_2|x),\mc{R}_1\times\mc{R}_2)$ is {\em no more 
degraded}. Thus, the capacity region expressions given for GBC can not be 
carried over to QBC.

Through simulations, we observed that in QBC, even in presence of a 
Gaussian noise with $\sigma_1^2<\sigma_2^2$, $I(x_2;r_1)$ is {\em not 
always greater} than $I(x_2;r_2)$. Table \ref{tab_mutual} shows a listing 
of the mutual information for a two-user QBC when both the users use a 
1-bit uniform quantizer and the input messages for both the users are 
from 4-QAM input alphabet at SNR1 = 10 dB and SNR2 = 7 dB. Observe that 
at $\alpha =0.6$ and $0.8$,  $I(x_2;r_1) < I(x_2;r_2)$.
Hence, user 1 can not decode user 2's message when $I(x_2;r_1) < I(x_2;r_2)$ 
and the rate of user 2's message is $I(x_2;r_2)$, which, in turn, implies 
that user 1 can not do successive decoding. However, if we set the rate of 
user 2 to $\min\{I(x_2;r_2), \; I(x_2;r_1)\}$, then it is guaranteed that 
both user 1 and user 2 can decode user 2's message and user 1 can do 
successive decoding.

\begin{table}[h]
\begin{center}
\begin{tabular}{|l|c|c|c|c|}
\hline
Mutual Information& $\alpha=0.2$ & $\alpha=0.4$ & $\alpha=0.6$ & $\alpha=0.8$ \\ \hline \hline
$I(x_1;r_1|x_2)$ & 0.08083 &0.37272 & 0.93188 & 1.59350 \\ \hline
$I(x_1;r_1)$ & 0.00893 & 0.15668 & 0.71584 & 1.52160 \\ \hline
$I(x_1;r_2)$ & 0.03572 & 0.20718 & 0.60551 & 1.19670 \\ \hline
$I(x_2;r_1)$ & 1.52160 & 0.71584 & {\bf 0.15668} & {\bf 0.00893} \\ \hline
$I(x_2;r_2)$ & 1.19670 & 0.60551 & {\bf 0.20718} & {\bf 0.03572} \\ \hline
$I(x_2;r_2|x_1)$ & 1.31920 & 0.82872 & 0.43039 & 0.15825 \\ \hline
\end{tabular}
\end{center}
\caption{\footnotesize{Mutual information for a two-user QBC when both the 
users use a 1-bit uniform quantizer and the input messages for both the 
users are from a 4-QAM alphabet at SNR1= 10 dB and SNR2 = 7 dB.}}
\label{tab_mutual}
\vspace{-5mm}
\end{table}

Based on the above observation, we now propose an achievable rate region 
for two-user QBC. We consider two schemes characterizing two different 
coding/decoding procedures to arrive at the proposed achievable rate 
region of QBC.

{{\bfseries Scheme 1}: \em  User 1 does successive decoding and user 2 
decodes its message alone.} 

User 1 can achieve a rate of $I(x_1;r_1\;|\;x_2)$ by successive decoding 
(i.e., user 1 will cancel the interference due to user 2's message and 
then it will decode its own message) only when it can decode user 2's 
message error free. From the observations made in Table \ref{tab_mutual}, 
we know that $I(x_2;r_1)$ is not always greater than $I(x_2;r_2)$ and 
hence, for user 1 to decode user 2's message error free, user 2's 
information must be restricted to a rate of 
$\min\{I(x_2;r_2), \; I(x_2;r_1)\}$. Thus, the set of achievable rate 
pairs $(R_1^{(1)},R_2^{(1)})$ when user 1 does successive decoding and 
user 2 decodes its message alone, is given by
\begin{eqnarray}
R_1^{(1)} &\leq& I(x_1;r_1\;|\;x_2) \label{eq_s1r1} \\
R_2^{(1)} &\leq& \min\{I(x_2;r_2), \; I(x_2;r_1)\}. \label{eq_s1r2}
\end{eqnarray}

{{\bfseries Scheme 2}: \em User 2 does successive decoding and user 1 
decodes its message alone.} 

Similarly, user 2 can achieve a rate of $I(x_2;r_2|x_2)$ by successive 
decoding only when the information to user 1 is restricted to a rate of 
$\min\{I(x_1;r_1), \; I(x_1;r_2)\}$. Thus, the set of achievable rate 
pairs $(R_1^{(2)},R_2^{(2)})$, when user 2 does successive decoding and 
user 1 decodes his message alone, is given by
\begin{eqnarray}
R_1^{(2)} &\leq& \min\{I(x_1;r_1), \; I(x_1;r_2)\} \label{eq_s2r1} \\
R_2^{(2)} &\leq& I(x_2;r_2\;|\;x_1). \label{eq_s2r2}
\end{eqnarray}

Since any line joining a pair of achievable rate pairs in the above two 
schemes is also achievable by {\em time sharing}, we propose the achievable 
rate region of QBC, $\mc{S}$, as the set of all rate pairs $(R_1,R_2)$ 
which are in the convex hull \cite{boyd} of the union of the achievable 
rate pairs of the above two schemes. The proposed achievable rate region, 
$\mc{S}$, is then given by

\vspace{-2mm}
{\small
\begin{equation}
\mc{S}=\{(R_1,R_2)\;|\; (R_1,R_2) \in conv(\;(R_1^{(1)},R_2^{(1)})\; \cup \;(R_1^{(2)},R_2^{(2)})) \}, \label{eq_ccr1r2}
\end{equation}
}

\vspace{-2mm}
where $conv(.)$ denotes convex hull, and $(R_1^{(1)},R_2^{(1)})$ satisfies 
(\ref{eq_s1r1}),(\ref{eq_s1r2}) and $(R_1^{(2)},R_2^{(2)})$ satisfies 
(\ref{eq_s2r1}),(\ref{eq_s2r2}).

\begin{figure*}
\centering
{\small
\begin{eqnarray}
&&\hsp{-19mm}R_1^{(1)} \;\; \leq \;\;\; \log_2(N_1)-\frac{1}{N_1N_2} \sum\limits_{k=1}^{2^{2b_1}} \sum\limits_{l_1=1}^{N_1} \sum\limits_{m_1=1}^{N_2}
p_{r_1|x_1,x_2}(\mc{R}_1(k)\;|\;\mc{X}_1(l_1),\mc{X}_2(m_1)) \nonumber\\
&&\hspace{6cm} \times \, \log_2\left\{\frac{\displaystyle \sum\nolimits_{l_2=1}^{N_1}p_{r_1|x_1,x_2}(\mc{R}_1(k)\;|\;\mc{X}_1(l_2),\mc{X}_2(m_1))}
{\displaystyle p_{r_1|x_1,x_2}(\mc{R}_1(k)\;|\;\mc{X}_1(l_1),\mc{X}_2(m_1))}\right\}. \label{eq_R11}
\end{eqnarray}
\hrulefill
\begin{eqnarray}
R_2^{(1)} & \leq & \min \left\{
\log_2(N_2)-\frac{1}{N_1N_2} \sum\limits_{k=1}^{2^{2b_1}} \sum\limits_{l_1=1}^{N_1} \sum\limits_{m_1=1}^{N_2}
p_{r_1|x_1,x_2}(\mc{R}_1(k)\;|\;\mc{X}_1(l_1),\mc{X}_2(m_1)) \right. \nonumber \\
&&\hspace{6cm} \times \, \log_2\left\{\frac{\displaystyle \sum\nolimits_{l_2=1}^{N_1} \sum\nolimits_{m_2=1}^{N_2} p_{r_1|x_1,x_2}(\mc{R}_1(k)\;|\; \mc{X}_1(l_2),\mc{X}_2(m_2))}
{\displaystyle \sum\nolimits_{l_3=1}^{N_1} p_{r_1|x_1,x_2}(\mc{R}_1(k)\;|\;\mc{X}_1(l_3),\mc{X}_2(m_1))}\right\}, \nn \\
&&\hsp{9mm} \log_2(N_2)-\frac{1}{N_1N_2} \sum\limits_{k=1}^{2^{2b_2}} \sum\limits_{l_1=1}^{N_1} \sum\limits_{m_1=1}^{N_2}
p_{r_2|x_1,x_2}(\mc{R}_2(k)\;|\;\mc{X}_1(l_1),\mc{X}_2(m_1)) \nonumber \\
&& \left. \hspace{6cm} \times \, \log_2\left\{\frac{\displaystyle \sum\nolimits_{l_2=1}^{N_1} \sum\nolimits_{m_2=1}^{N_2} p_{r_2|x_1,x_2}(\mc{R}_2(k)\;|\; \mc{X}_1(l_2),\mc{X}_2(m_2))}
{\displaystyle \sum\nolimits_{l_3=1}^{N_1} p_{r_2|x_1,x_2}(\mc{R}_2(k)\;|\;\mc{X}_1(l_3),\mc{X}_2(m_1))} \right\} \right\}.
\label{eq_R21}
\end{eqnarray}
\hrulefill
\begin{eqnarray}
R_1^{(2)} & \leq & \min \left\{
\log_2(N_1)-\frac{1}{N_1N_2} \sum\limits_{k=1}^{2^{2b_1}} \sum\limits_{l_1=1}^{N_1} \sum\limits_{m_1=1}^{N_2}
p_{r_1|x_1,x_2}(\mc{R}_1(k)\;|\;\mc{X}_1(l_1),\mc{X}_2(m_1)) \right. \nonumber \\
&&\hspace{6cm} \times \, \log_2\left\{\frac{\displaystyle \sum\nolimits_{l_2=1}^{N_1} \sum\nolimits_{m_2=1}^{N_2} p_{r_1|x_1,x_2}(\mc{R}_1(k)\;|\; \mc{X}_1(l_2),\mc{X}_2(m_2))}
{\displaystyle \sum\nolimits_{m_3=1}^{N_2} p_{r_1|x_1,x_2}(\mc{R}_1(k)\;|\;\mc{X}_1(l_1),\mc{X}_2(m_3))}\right\}, \nn \\
&&\hsp{9mm} \log_2(N_1)-\frac{1}{N_1N_2} \sum\limits_{k=1}^{2^{2b_2}} \sum\limits_{l_1=1}^{N_1} \sum\limits_{m_1=1}^{N_2}
p_{r_2|x_1,x_2}(\mc{R}_2(k)\;|\;\mc{X}_1(l_1),\mc{X}_2(m_1)) \nonumber \\
&& \left. \hspace{6cm} \times \, \log_2\left\{\frac{\displaystyle \sum\nolimits_{l_2=1}^{N_1} \sum\nolimits_{m_2=1}^{N_2} p_{r_2|x_1,x_2}(\mc{R}_2(k)\;|\; \mc{X}_1(l_2),\mc{X}_2(m_2))}
{\displaystyle \sum\nolimits_{m_3=1}^{N_2} p_{r_2|x_1,x_2}(\mc{R}_2(k)\;|\;\mc{X}_1(l_1),\mc{X}_2(m_3))} \right\} \right\}.
\label{eq_R12}
\end{eqnarray}
\hrulefill
\begin{eqnarray}
&&\hsp{-19mm}R_2^{(2)} \;\; \leq \;\;\;\log_2(N_2)-\frac{1}{N_1N_2} \sum\limits_{k=1}^{2^{2b_2}} \sum\limits_{l_1=1}^{N_1} \sum\limits_{m_1=1}^{N_2}
p_{r_2|x_1,x_2}(\mc{R}_2(k)\;|\;\mc{X}_1(l_1),\mc{X}_2(m_1)) \nonumber\\
&&\hspace{6cm} \times \, \log_2\left\{\frac{\displaystyle \sum\nolimits_{m_2=1}^{N_2}p_{r_2|x_1,x_2}(\mc{R}_2(k)\;|\;\mc{X}_1(l_1),\mc{X}_2(m_2))}
{\displaystyle p_{r_2|x_1,x_2}(\mc{R}_2(k)\;|\;\mc{X}_1(l_1),\mc{X}_2(m_1))}\right\}. \label{eq_R22}
\end{eqnarray}
\hrulefill
\vspace{-6mm}
}
\end{figure*}

The mutual information in the expressions (\ref{eq_s1r1}), (\ref{eq_s1r2}), (\ref{eq_s2r1}), (\ref{eq_s2r2}) are calculated using the probability distribution
\begin{eqnarray}
&& \hsp{-5mm}p\left(r_1=\mc{R}_1(k) \, | \, x_1=\mc{X}_1(l),x_2=\mc{X}_2(m)\right)
 \nonumber \\
&& \hsp{-5mm}= p(r_1^I=\mc{R}_1^I(k), \, r_1^Q=\mc{R}_1^Q(k) \, | \, x_1=\mc{X}_1(l),x_2=\mc{X}_2(m)) \nonumber \vspace{1mm}\\
&& \hsp{-5mm}= p(z_1^I \in \mc{F}_1(\mc{X}_1^I(l),\mc{X}_2^I(m),\mc{R}_1^I(k))) \nonumber \\
&& \hspace{15mm} \times \, p(z_1^Q \in \mc{F}_1(\mc{X}_1^Q(l),\mc{X}_2^Q(m),\mc{R}_1^Q(k))),
\label{eq:pr112}
\end{eqnarray}
where $j=\sqrt{-1}$,  and $\mc{R}_1(i)$, $\mc{X}_1(i)$ and $\mc{X}_2(i)$
refer to the $i$th element of sets $\mc{R}_1$, $\mc{X}_1$ and $\mc{X}_2$,
respectively.
The region $\mc{F}_1(.)$ is defined as
\begin{eqnarray}
\mc{F}_1(p,q,t)=\{n \in \mathbb{R} \, | \, \mbf{Q}_{b_1}(p+q+n)=t\},
\end{eqnarray}
and $n \sim \mc{N}(0,\sigma_1^2/2)$.
From (\ref{eq:pr112}), the marginal probability distributions 
$p(r_1|x_1)$, $p(r_1|x_2)$ and $p(r_1)$ are calculated as
\begin{eqnarray}
&& \hsp{-10mm}p(r_1=\mc{R}_1(k) \, | x_1=\mc{X}_1(l))\nonumber \\
&& \hsp{-10mm}= \frac{1}{N_2}\sum_{m=1}^{N_2}p(r_1=\mc{R}_1(k) \, |\, x_1=\mc{X}_1(l),x_2=\mc{X}_2(m)), \label{eq:pr11}
\end{eqnarray}
\vspace{-5mm}
\begin{eqnarray}
&& \hsp{-10mm}p(r_1=\mc{R}_1(k) \, | x_2=\mc{X}_2(m))\nonumber \\
&& \hsp{-10mm}= \frac{1}{N_1}\sum_{l=1}^{N_1}p(r_1=\mc{R}_1(k) \, |\, x_1=\mc{X}_1(l),x_2=\mc{X}_2(m)), \label{eq:pr12}
\end{eqnarray}
\vspace{-5mm}
\begin{eqnarray}
&& \hsp{-28mm} p(r_1=\mc{R}_1(k)) \nonumber \\
&& \hsp{-28mm} = \frac{1}{N_2}\sum_{m=1}^{N_2}p(r_1=\mc{R}_1(k)\,|\,x_2=\mc{X}_2(m)).
\label{eq:pr1}
\end{eqnarray}
Similarly, the probability distributions $p(r_2|x_1, x_2)$, $p(r_2|x_1)$, 
$p(r_2|x_2)$ and $p(r_2)$ can be calculated. Using the above probability 
distributions, the final expressions of (\ref{eq_s1r1})-(\ref{eq_s2r2}) 
are given by Eqns. (\ref{eq_R11})-(\ref{eq_R22}), which are listed above.

In the illustration of numerical results, we plot the boundary of the 
achievable rate region of two-user QBC by varying the proportion of 
power ($\alpha$) allocated to each user from $0$ to $1$ and finding the 
achievable rate pairs using (\ref{eq_ccr1r2}). When both $x_1$ and $x_2$ 
take values from the same signal set, we consider rotation of the second 
user's signal set by an angle $\theta$ with respect to the first user's 
signal set for further enlargement of the achievable rate region, i.e., 
\begin{eqnarray}
\mc{X}_2 \,\, \Define \,\, \{u \, e^{j\theta}\; | \; u \in \mc{X}_1\},
\label{eq:chi2}
\end{eqnarray}
where $\theta$ is the rotation angle. We observe that, the rate expressions 
now become a function of $\theta$, and hence they are explicitly denoted 
as $R^{(1)}_1(\theta)$, $R^{(1)}_2(\theta)$, $R^{(2)}_1(\theta)$ and 
$R^{(2)}_2(\theta)$. The achievable rate region of QBC with rotation, 
$\mc{S}_{\theta}$, is then given by

\vspace{-3mm}
{\small
\begin{eqnarray}
&&\hspace{-12mm}\mc{S}_{\theta}=\Big\{(R_1,R_2)\; | \; (R_1,R_2) \in  \nn\\ &&\hspace{-10mm}conv\Big(\;\hsp{-1mm}\bigcup\limits_{\theta \in (0, 2\pi)} \hsp{-2mm}\{(R_1^{(1)}(\theta),R_2^{(1)}(\theta))\; \cup \;(R_1^{(2)}(\theta),R_2^{(2)}(\theta))\}\Big)\Big\}. \label{eq:cctr1r2}
\end{eqnarray}
}

\vspace{-2mm}
\subsection{QBC with Uniform Quantizer} 
\label{sec_uni}
In this subsection, we study the achievable rate region of two-user QBC 
with uniform receiver quantization.

\subsubsection{Uniform Quantizer}
A uniform $b$-bit quantizer, $\mbf{Q}_b(.)$ acting on the
real component of the analog received signal $y$ is given by
\begin{eqnarray}
\mbf{Q}_b(y^I) &\hspace{-2mm} \Define & \hspace{-2mm} \left\{\begin{array}{cl}
+1, & \zeta(y^I) > (2^{b-1}-1)\\
-1, &  \zeta(y^I) < -(2^{b-1}-1)\\
\frac{\displaystyle2\zeta(y^I)+1}{\displaystyle2^b-1}, & \mbox{otherwise},
\end{array} \right.
\label{eq:unir}
\end{eqnarray}
where
$\zeta(y^I)\Define \left\lfloor\frac{y^I}{X^I}\frac{(2^b-1)}{2} \right\rfloor$ 
and $X^I$ is defined in (\ref{eq_xiq}).
Similarly, 
\begin{eqnarray}
\mbf{Q}_b(y^Q) & \hspace{-2mm} \Define & \hspace{-2mm} \left\{\begin{array}{cl}
+1, & \zeta(y^Q) > (2^{b-1}-1)\\
-1, &  \zeta(y^Q) < -(2^{b-1}-1)\\
\frac{\displaystyle2\zeta(y^Q)+1}{\displaystyle2^b-1}, & \mbox{otherwise},
\end{array} \right.
\label{eq:unii}
\end{eqnarray}
where $\zeta(y^Q) \Define \left\lfloor \frac{y^Q}{X^Q}\frac{(2^b-1)}{2} \right\rfloor$ 
and $X^Q$ is defined in (\ref{eq_xiq}).

We assume that the user 1 uses a $b_1$-bit uniform quantizer and user 2 
uses a $b_2$-bit uniform quantizer. Applying the above uniform quantizer 
to the analog received signal at the users 1 and 2, their quantized 
outputs on the real and imaginary components are given by
\begin{eqnarray}
r_1^I=Q_{b_1}(y_1^I), && r_1^Q=Q_{b_1}(y_1^Q), \label{eq_r1qb}\\
r_2^I=Q_{b_2}(y_2^I), && r_2^Q=Q_{b_2}(y_2^Q). \label{eq_r2qb}
\end{eqnarray}
With the uniform quantizer defined in (\ref{eq_r1qb}) and (\ref{eq_r2qb}), 
we numerically evaluate the proposed achievable rate region of two-user 
QBC using (\ref{eq:cctr1r2}) or (\ref{eq_ccr1r2}), the results of which are 
discussed in the following subsection. 

\begin{figure}
\begin{center}
\subfigure[SNR1 = 13 dB, SNR2 = 15 dB]{\label{fig_13_15} \includegraphics[scale=0.35]{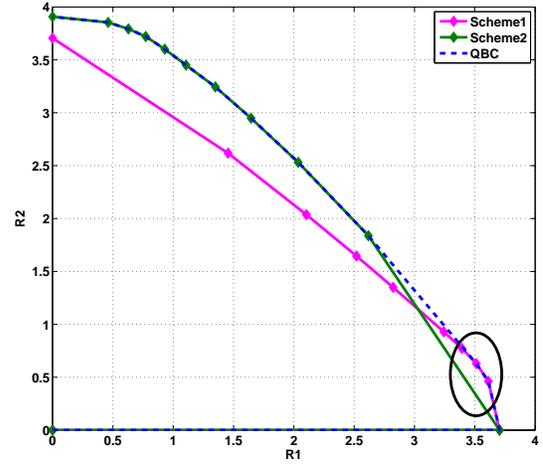}}
\subfigure[SNR1 = 15 dB, SNR2 = 13 dB]{\;\label{fig_15_13}
\includegraphics[scale=0.35]{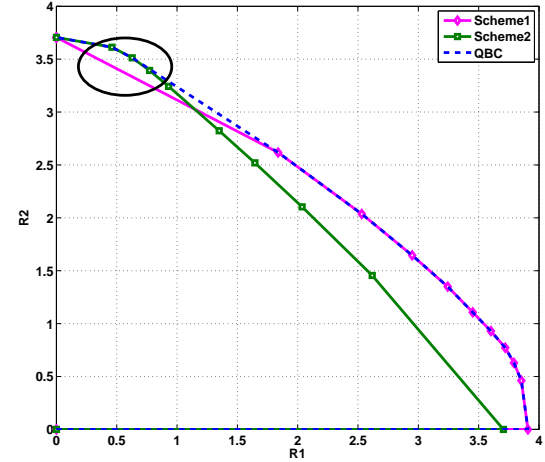}}
\subfigure[SNR1 = 15 dB, SNR2 = 15 dB]{\label{fig_15_15}
\includegraphics[scale=0.35]{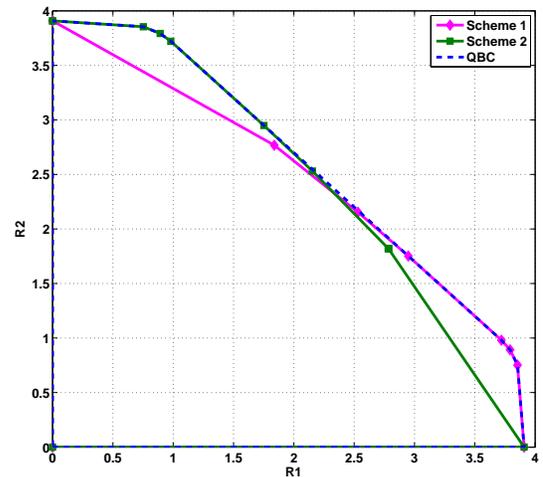}}
\caption{Plots of the boundary of the proposed achievable rate region of 
two-user QBC when the input alphabet for user 1 is 16-QAM and the input 
alphabet for user 2 is a rotated 16-QAM with different SNR combinations 
at the two users. The users use $b_1=b_2=4$-bit uniform receiver quantizer.} 
\label{fig_snr_comp}
\end{center}
\vspace{-5mm}
\end{figure}

\subsubsection{Results and Discussion}
In Fig. \ref{fig_snr_comp}, we first illustrate the significance of using 
the two schemes instead of assuming that the user with high SNR alone does 
successive decoding. Figures \ref{fig_13_15}, \ref{fig_15_13} and 
\ref{fig_15_15} show the proposed achievable rate region of two-user QBC 
when the input alphabet for user 1 is 16-QAM and the input alphabet for 
user 2 is a rotated 16-QAM, and both the users use 4-bit uniform receiver 
quantization. In Fig. \ref{fig_13_15}, SNR1 = 13 dB and SNR2 = 15 dB. In 
Fig. \ref{fig_15_13}, SNR1 = 15 dB, SNR2 = 13 dB, and in Fig. \ref{fig_15_15}, 
SNR1 = SNR2 = 15 dB. We observe that most of the contribution to the proposed 
achievable rate region of QBC is due to the scheme of the user with high SNR 
doing successive decoding and the user with low SNR decoding his message 
alone. For example, observe the performance of scheme 2 in Fig. 
\ref{fig_13_15} and scheme 1 in Fig. \ref{fig_15_13}. However, there is an 
appreciable contribution to the proposed achievable rate region of QBC when 
the user with low SNR performs successive decoding and the user with high 
SNR  decodes his message alone, especially when the proportion of the total
transmit power allotted to that user (the one with low SNR) is more than 
that of the other. For instance, observe the performance in the {\em circled 
regions} of scheme 1 in Fig. \ref{fig_13_15} and scheme 2 in Fig. 
\ref{fig_15_13}. When the SNR of both the users are same, equal contribution 
is made by the two schemes to the proposed achievable rate region  of QBC, 
which is illustrated in Fig. \ref{fig_15_15}.

\begin{figure}[h]
\vspace{-4mm}
\hspace{-3mm}
\epsfysize=7.8cm
\epsfxsize=9.4cm
\epsfbox{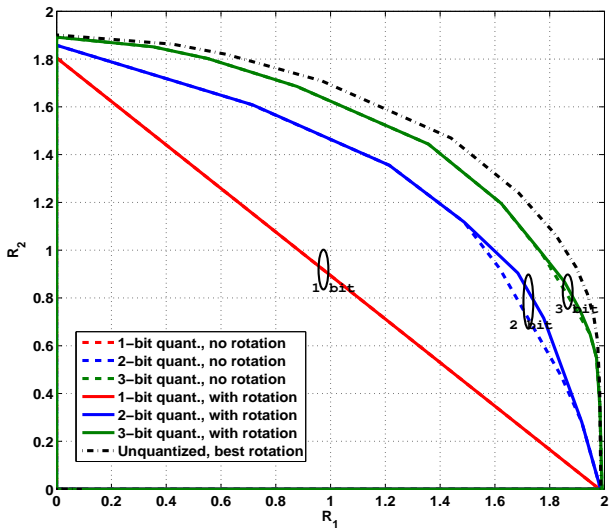}
\vspace{-10mm}
\caption{Comparison of the proposed achievable rate region of two-user QBC 
when both the users use uniform quantizer of same resolution i.e., $b_1=b_2$ 
at SNR1 = 10 dB and SNR2 = 7 dB. The input alphabet for user 1 is 4-QAM and 
the input alphabet for user 2 is a rotated 4-QAM.
}
\label{fig_1234uni}
\vspace{-2mm}
\end{figure}

Figure \ref{fig_1234uni} shows the significance of rotation on the proposed 
achievable rate region of QBC when both the users use uniform quantizer of 
same resolution, i.e., $b_1=b_2$ at SNR1 = 10 dB and SNR2 = 7 dB. The input 
alphabet for user 1 is 4-QAM and the input alphabet for user 2 is a rotated 
4-QAM. We observe that there is no increase in the achievable rate region 
for a 1-bit uniform quantizer due to rotation compared to that of the 
achievable rate region without rotation. For $b_1=b_2=2$ or $3$ bit uniform 
quantizers, there is a small increase in the achievable rate region due to 
rotation compared to the achievable rate region without rotation only when 
$\alpha$ is around 0.5. The reason could be that rotation gives significant 
enlargement in the achievable rate region only when the sum signal set is 
not uniquely decodable, i.e., when there is no one-to-one correspondence 
between the elements in the set $\mc{X}$ to the elements in the set 
$\mc{X}_1 \times \mc{X}_2$. This happens more only when $\alpha$ is 
around 0.5. For instance, when $\alpha=0.5$, $\mc{X}_1=\mc{X}_2$ and 
thus the set $\mc{X}$ is not uniquely decodable. Hence, when $\alpha=0.5$, 
rotation by even a small angle makes the set $\mc{X}$ to be uniquely 
decodable resulting in an increase in the achievable rate region of QBC.
Finally, we have computed the proposed achievable rate region for QBC 
with asymmetric quantizers also, i.e., with $b_1 \neq b_2$.

\section{Quantized Multiple Access Channel}
\label{sec_qmac}
In this section, we study the achievable rate region of two-user QMAC
\cite{qmac}.

\subsection{System Model}
\label{sec2a}
\begin{figure}[t]
\centering
\epsfysize=7.0cm
\epsfxsize=7.50cm
\hspace{-4mm}
\epsfbox{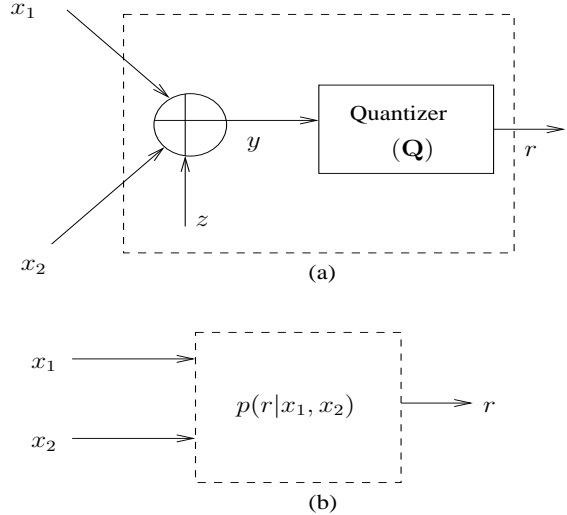}
\vspace{-2.5mm}
\caption{(a) Two-user Gaussian MAC model with quantized output. 
(b) Equivalent DMC.}
\label{fig:MAC}
\vspace{-2.0mm}
\end{figure}

Consider a two-user Gaussian MAC channel. Let $x_1$ and $x_2$ be the
symbols transmitted by the first and second user, respectively. Let
$x_1 \in \mathcal{X}_1$ and $x_2 \in \mathcal{X}_2$, where $\mc{X}_1$
and $\mc{X}_2$ are finite signal sets with $N_1$ and $N_2$ equi-probable 
complex entries, respectively. Let $z \sim \mc{CN}(0, \sigma^2)$ be the 
additive white Gaussian noise at the receiver. The analog received signal 
is then given by
\begin{eqnarray}
y=x_1+x_2+z. \label{eq:y}
\end{eqnarray}
The analog received signal, $y$, is quantized by a complex quantizer 
$\mathbf{Q}$, resulting in the output $r$, as shown in Fig. \ref{fig:MAC}.

The quantizer $\mathbf{Q}$ is composed of two similar quantizers
acting independently on the real and imaginary components of the received
analog signal, $y$. The real and imaginary components of the quantized
output $r$ are then given by
\begin{equation}
r^{I}\,=\,\mathbf{Q}_b(y^I), \quad r^{Q}\,=\,\mathbf{Q}_b(y^Q),
\label{eq:rirq}
\end{equation}
where the function $\mbf{Q}_b(.)$ models a receiver quantizer having a
resolution of $b$ bits. $\mbf{Q}_b(.)$ is a mapping from the set of real
numbers $\mathbb{R}$ to a finite alphabet set $\mc{S}_b$ of cardinality
$2^b$, i.e.,
\begin{equation}
\mbf{Q}_b: \mathbb{R} \mapsto \mc{S}_b, \quad \mc{S}_b \subset \mathbb{R},
\quad  |\mc{S}_b|=2^b.
\label{eq:Q}
\end{equation}
Let $\mc{R}$ be defined as
\begin{equation}
\mc{R}=\{c+jd \;|\; c,d \in \mc{S}_b\}, \; |\mc{R}|=2^{2b}, \; j=\sqrt{-1}. 
\label{eq:setr}
\end{equation}
Thus the quantized output, $r$, takes values from the set $\mc{R}$.
Henceforth, we refer to the above system model as {\em quantized MAC
(QMAC)}.

\subsection{Achievable Rate Region of QMAC}
In this subsection, we derive analytical expressions for the rate region 
of a two-user QMAC. From the Fig. \ref{fig:MAC}, we observe that the 
effective multiple-access channel after receiver quantization,
$(\mc{X}_1\times\mc{X}_2, \, p(r|x_1,x_2), \, \mc{R})$, is a discrete 
memoryless channel (DMC)  with the transition probabilities derived and 
given in (\ref{eq:pr12}). Let $R_1$ and $R_2$ represent the rates achieved 
by user 1 and user 2, respectively. Since QMAC is a discrete memoryless 
multiple-access channel, the achievable rate region \cite{cover99} is 
the set of all rate pairs $(R_1, R_2)$ satisfying
\begin{eqnarray}
 R_1 & \leq & I(x_1;r \, | \, x_2) \label{eq:r1}\\
 R_2 & \leq & I(x_2;r \, | \, x_1) \label{eq:r2}\\
 R_1+R_2 & \leq & I(x_1, x_2;r ) \nonumber \\
& = & I(x_2;r)+I(x_1;r \, | \, x_2).
\end{eqnarray}
The mutual information $I(x_2;r)$, $I(x_1;r|x_2)$ are given by
\begin{eqnarray}
I(x_2;r) & = & H(r)-H(r|x_2) \label{eq:I2r}\\
I(x_1;r|x_2) & = & H(r|x_2)-H(r|x_1,x_2),
\label{eq:I1r2}
\end{eqnarray}
where the entropies in (\ref{eq:I2r}) and (\ref{eq:I1r2}) are calculated
using the probability distribution function
\begin{eqnarray}
&&\hspace{-5mm} p\left(r=\mc{R}(k) \, | \, x_1=\mc{X}_1(l),x_2=\mc{X}_2(m)\right)
\nonumber \\
&& \hspace{-4mm} =  \hspace{-0cm} p(r^I=\mc{R}^I(k), \, r^Q=\mc{R}^Q(k) \, | \, x_1=\mc{X}_1(l),x_2=\mc{X}_2(m)) \nonumber \vspace{1mm}\\
&& \hspace{-4mm} =  \hspace{-0cm} p(z^I \in \mc{F}(\mc{X}_1^I(l),\mc{X}_2^I(m),\mc{R}^I(k))) \nonumber \\
&& \hspace{15mm} \times \, p(z^Q \in \mc{F}(\mc{X}_1^Q(l),\mc{X}_2^Q(m),\mc{R}^Q(k))),
\label{eq:pr12}
\end{eqnarray}
where $j=\sqrt{-1}$,  and $\mc{S}_b(i)$, $\mc{X}_1(i)$ and $\mc{X}_2(i)$
refer to the $i$th element of sets $\mc{S}_b$, $\mc{X}_1$ and $\mc{X}_2$,
respectively.
$\mc{R}^I(k),\mc{X}_1^{I}(l)$, $\mc{X}_1^{Q}(l)$ and
$\mc{R}^Q(k),\mc{X}_2^{I}(m)$, $\mc{X}_2^{Q}(m)$ are
the real and imaginary parts of $\mc{R}(k),\mc{X}_1(l)$ and
$\mc{X}_2(m)$, respectively.

The region $\mc{F}(.)$ is defined as
\begin{equation}
\mc{F}(p,q,t)=\{n \in \mathbb{R} \, | \, \mbf{Q}_b(p+q+n)=t\},
\end{equation}
and $n \sim \mc{N}(0,\sigma^2/2)$.
From (\ref{eq:pr12}), the probability distributions $p(r|x_2)$ and $p(r)$
are calculated as
\begin{eqnarray}
&& \hspace{-21mm} p(r=\mc{R}(k) \, | x_2=\mc{X}_2(m)) \nonumber \\
&& \hspace{-21mm} =  \hspace{-0cm} \frac{1}{N_1}\hspace{-0.5mm}\sum_{l=1}^{N_1}p(r=\mc{R}(k)|x_1=\mc{X}_1(l),x_2=\mc{X}_2(m));
\label{eq:pr2}
\end{eqnarray}
\begin{eqnarray}
&&\hspace{-32mm}
p(r=\mc{R}(k))  \nonumber \\
&& \hspace{-32mm} =  \hspace{-0cm} \frac{1}{N_2}\sum_{m=1}^{N_2}p(r=\mc{R}(k)|x_2=\mc{X}_2(m)).
\label{eq:pr}
\end{eqnarray}

On substituting (\ref{eq:pr12}), (\ref{eq:pr2}), (\ref{eq:pr}) into
(\ref{eq:I2r}) and (\ref{eq:I1r2}), $I(x_2;r)$ and $I(x_1;r|x_2)$
can be computed. By symmetry, $I(x_1;r)$ and $I(x_2;r|x_1)$ can
be computed in a similar manner. The final expressions for the achievable
rate pairs $(R_1,R_2)$ are then given by (\ref{eq:R1}), (\ref{eq:R2}),
and (\ref{eq:R12}), presented on the top of the next page.

\begin{figure*}
\centering
\begin{eqnarray}
R_1 & \leq & \log_2(N_1)-\frac{1}{N_1N_2} \sum\limits_{k=1}^{2^{2b}} \sum\limits_{l_1=1}^{N_1} \sum\limits_{m_1=1}^{N_2}
p_{r|x_1,x_2}(\mc{R}(k)\;|\;\mc{X}_1(l_1),\mc{X}_2(m_1)) \nonumber\\
&&\hspace{6cm} \times \, \log_2\left\{\frac{\displaystyle \sum\nolimits_{l_2=1}^{N_1}p_{r|x_1,x_2}(\mc{R}(k)\;|\;\mc{X}_1(l_2),\mc{X}_2(m_1))}
{\displaystyle p_{r|x_1,x_2}(\mc{R}(k)\;|\;\mc{X}_1(l_1),\mc{X}_2(m_1))}\right\} \label{eq:R1}
\end{eqnarray}
\hrulefill
\begin{eqnarray}
R_2 & \leq & \log_2(N_2)-\frac{1}{N_1N_2} \sum\limits_{k=1}^{2^{2b}} \sum\limits_{l_1=1}^{N_1} \sum\limits_{m_1=1}^{N_2}
p_{r|x_1,x_2}(\mc{R}(k)\;|\;\mc{X}_1(l_1),\mc{X}_2(m_1)) \nonumber \\
&&\hspace{6cm} \times \, \log_2\left\{\frac{\displaystyle \sum\nolimits_{m_2=1}^{N_2}p_{r|x_1,x_2}(\mc{R}(k)\;|\; \mc{X}_1(l_1),\mc{X}_2(m_2))}
{\displaystyle p_{r|x_1,x_2}(\mc{R}(k)\;|\;\mc{X}_1(l_1),\mc{X}_2(m_1))}\right\} \label{eq:R2}
\end{eqnarray}
\hrulefill
\begin{eqnarray}
R_1+R_2 & \leq & \log_2(N_1N_2)-\frac{1}{N_1N_2} \sum\limits_{k=1}^{2^{2b}} \sum\limits_{l_1=1}^{N_1} \sum\limits_{m_1=1}^{N_2}
p_{r|x_1,x_2}(\mc{R}(k)\;|\;\mc{X}_1(l_1),\mc{X}_2(m_1)) \nonumber \\
&&\hspace{4cm}\times \, \log_2 \left\{ \frac{\displaystyle \sum\nolimits_{l_2=1}^{N_1}\sum\nolimits_{m_2=1}^{N_2}p_{r|x_1,x_2}(\mc{R}(k)\;|\; \mc{X}_1(l_2),\mc{X}_2(m_2))}{\displaystyle p_{r|x_1,x_2}(\mc{R}(k)\;|\;\mc{X}_1(l_1),\mc{X}_2(m_1))} \right\} \label{eq:R12}
\end{eqnarray}
\hrulefill
\vspace{-4.5mm}
\end{figure*}

Now, let $\mathbb{A}_M \Define \{ -(M-1),\cdots,-1,\,1,\cdots,(M-1)\}$
be the $M$-PAM signal set, and
$\mathbb{A}_{M^2} \Define \{u+jv \; | \; u,v \in \mathbb{A}_M\}$
denote the corresponding $M^2$-QAM signal set.
We restrict the input of the first user to be from $M^2$-QAM
alphabet, and the second user input to be from a rotated version of the
first user's input alphabet, i.e., $\mc{X}_1=\mathbb{A}_{M^2}$, and
\begin{equation}
\mc{X}_2 \Define \{u \, e^{j\theta}\; | \; u \in \mc{X}_1\},
\label{eq:chi2}
\end{equation}
where $\theta$ is the rotation angle. We are interested in maximizing the 
sum rate $(R_1+R_2)$ achieved using the input alphabets $\mc{X}_1$ and 
$\mc{X}_2$ defined above. Since $R_1$ and $R_2$ are functions of $\theta$, 
we denote them by $R_1(\theta)$ and $R_2(\theta)$, respectively. For a 
given $b$-bit quantizer, the optimal rotation angle, $\theta^{opt}$, which 
maximizes the sum rate is given by
\begin{equation}
\theta^{opt}=\arg\max_{\{\theta | \mc{X}_1 \in \mathbb{A}_{M^2}, \mc{X}_2 \in \{ue^{j\theta} | u \in \mc{X}_1\}\}} R_1(\theta)+R_2(\theta).
\label{eq:thetaopt}
\end{equation}
In all the numerical results reported, the resolution of $\theta$ in the 
above optimization is set to $1^{\circ}$.

\subsection{QMAC with Uniform Quantizer}
\label{sec:uni}
In this subsection, we study the achievable two-user QMAC rate region
with a uniform $b$-bit quantizer. First, define the sum signal set as
\begin{equation}
\mc{X}_{sum}\,=\,\left\{x_1+x_2\, |\, x_1 \in \mc{X}_1,\, x_2 \in \mc{X}_2\right\}.
\end{equation}
Let $X^I$ and $X^Q$ be defined as
\begin{equation}
X^I \Define \max_{a \in \mc{X}_{sum}} |a^I|, \quad
X^Q \Define \max_{a \in \mc{X}_{sum}} |a^Q|.
\label{eq:xiq}
\end{equation}
Now, the function $\mbf{Q}_b(.)$ for the uniform $b$-bit quantizer on the
real component of the received signal $y$ is given by
\begin{equation}
\hspace{-2mm}
r^I=\mbf{Q}_b(y^I)\Define\left\{\begin{array}{cl}
+1, & \zeta(y^I) > (2^{b-1}-1)\\
-1, &  \zeta(y^I) < -(2^{b-1}-1)\\
\frac{\displaystyle2\zeta(y^I)+1}{\displaystyle2^b-1}, & \mbox{otherwise},
\end{array} \right.
\label{eq:unir}
\end{equation}
where
$\zeta(y^I)\Define \left\lfloor\frac{y^I}{X^I}\frac{(2^b-1)}{2} \right\rfloor$.
Similarly, for the imaginary component of $y$,
\begin{equation}
r^Q=\mbf{Q}_b(y^Q)\Define\left\{\begin{array}{cl}
+1, & \zeta(y^Q) > (2^{b-1}-1)\\
-1, &  \zeta(y^Q) < -(2^{b-1}-1)\\
\frac{\displaystyle2\zeta(y^Q)+1}{\displaystyle2^b-1}, & \mbox{otherwise},
\end{array} \right.
\label{eq:unii}
\end{equation}
where $\zeta(y^Q) \Define \left\lfloor \frac{y^Q}{X^Q}\frac{(2^b-1)}{2} \right\rfloor$.

With the uniform quantizer defined in (\ref{eq:unir}) and (\ref{eq:unii}),
we have numerically evaluated the rate region using (\ref{eq:R1}),
(\ref{eq:R2}) and (\ref{eq:R12}), the results of which are discussed next.

\begin{figure}
\hspace{-4mm}
\epsfysize=7.8cm
\epsfxsize=9.4cm
\hspace{-0mm}
\epsfbox{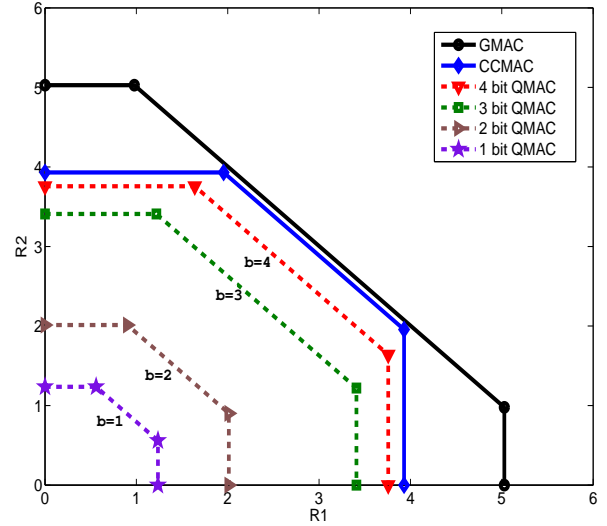}
\vspace{-8.0mm}
\caption{Rate region of two-user QMAC with uniform quantizer. User 1
transmits from 16-QAM signal set, and User 2 transmits from an optimally
rotated version of the first user's signal set. SNR per user = 15 dB.}
\label{fig:unirate}
\vspace{-6.0mm}
\end{figure}

\subsubsection{Results and Discussion}
\begin{figure}[h]
\centering
\subfigure[2-user sum signal set]{\includegraphics[scale=0.45]{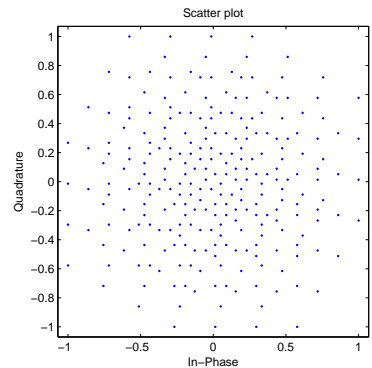}}
\subfigure[3-user sum signal set]{\includegraphics[scale=0.45]{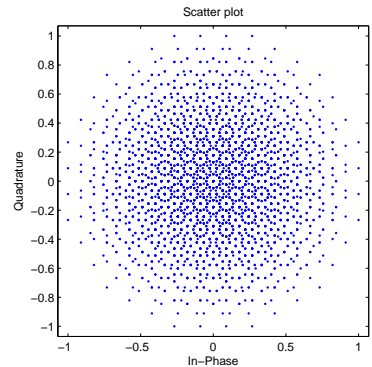}}
\caption{Scatter plots of two-user and three-user sum signal sets.}
\label{fig:scat}
\vspace{-6mm}
\end{figure}

In Fig. \ref{fig:unirate}, we plot the rate region of a two-user QMAC as a
function of the quantizer resolution, $b$, with User 1 using a 16-QAM input
alphabet and User 2 using an optimally rotated version of the 16-QAM alphabet,
as per (\ref{eq:thetaopt}), at SNR per user = 15 dB. The rate regions of GMAC
(Gaussian MAC with Gaussian inputs and no output quantization) and CCMAC
(Gaussian MAC with finite input and no output quantization \cite{harsh08})
are also plotted. From Fig. \ref{fig:unirate}, we observe that with low 
precision ADCs ($b=1$ or $2$ bits), the max. sum rate achieved with uniform 
receiver quantization is very poor compared to the max. sum rate of CCMAC. 
For instance, with a 2-bit uniform quantizer, the max. sum rate is 2.9144 bits
which is just 49.5\% of the max. sum rate of CCMAC (5.886 bits). To achieve a
max. sum rate close to CCMAC, increased quantization resolution is required.
For a fixed quantization resolution of $b$ bits, the degradation in the
rate region due to a uniform quantizer compared to CCMAC is expected to 
be more with increasing number of users. This is because the sum 
constellation becomes more and more dense around the origin, as 
illustrated in Fig. \ref{fig:scat}. Figure \ref{fig:scat}(a) shows the 
two-user sum signal set with User 1 using 16-QAM with no rotation and 
User 2 using 16-QAM with $45^{\circ}$ rotation. Figure \ref{fig:scat}(b) 
shows the three-user sum signal set; User 1 using 16-QAM with no rotation 
and Users 2 and 3 using 16-QAM with $30^{\circ}$ and $60^{\circ}$ rotations, 
respectively. It can be seen that the scatter plot for the three-user sum 
signal set is clustered more around the origin than that for the two-user 
sum signal set.

\begin{figure*}
\centering
\subfigure[3-bit {\em uniform} quantizer]{\epsfysize=4.0cm
\epsfxsize=15.00cm
\epsfbox{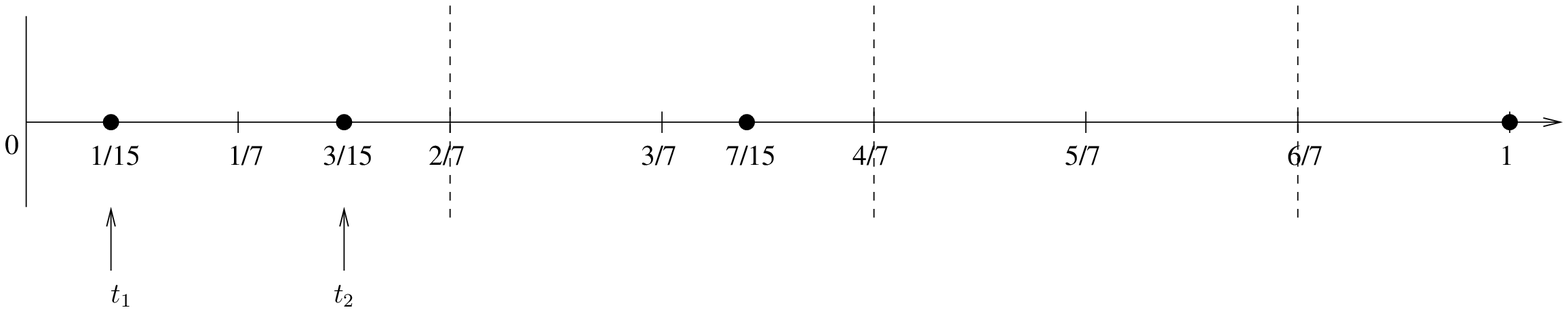}}
\subfigure[3-bit {\em non-uniform} quantizer]{\epsfysize=4.0cm
\epsfxsize=15.00cm
\epsfbox{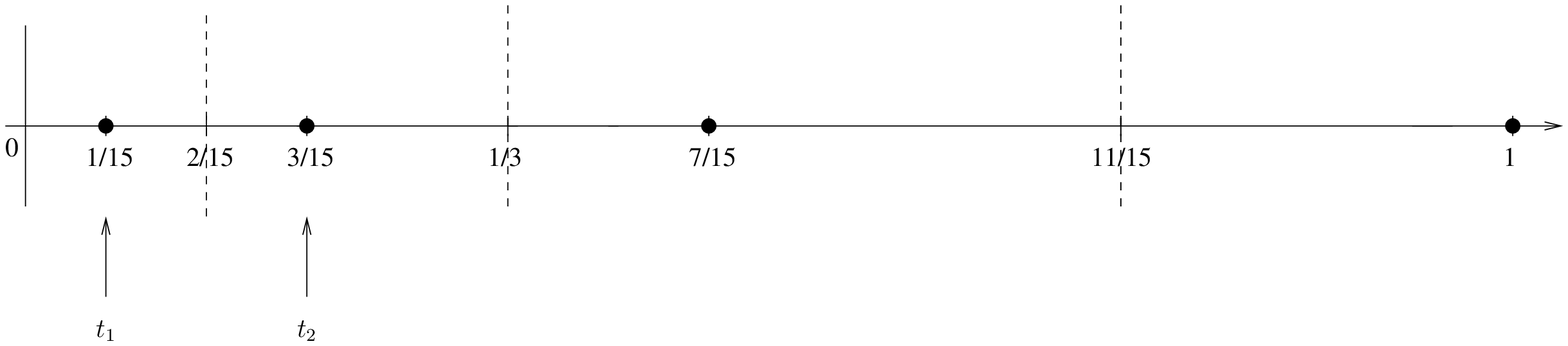}}
\caption{Plot of the quantization intervals for 3-bit uniform and
non-uniform quantizers. The boundaries between the quantization regions
are demarcated by the dotted lines
and the black dots show the points given in (\ref{eq:points}).
Since the quantizer is symmetric about the origin, only the positive
side is shown.
}
\label{fig:quan}
\vspace{-2mm}
\end{figure*}

\subsubsection{Motivation for a Non-uniform Quantizer}
Since the symbols in the sum signal set are densely distributed around the
origin, for a given quantization resolution of $b$ bits, the uniform
quantizer may not be the best quantizer in terms of the achievable
rate region. We highlight this point through a simple example.
Let
\begin{equation}
\left(\frac{x_1^I+x_2^I}{X^I}\right) \in \left\{-1,\frac{-7}{15},\frac{-3}{15},\frac{-1}{15},\frac{1}{15},\frac{3}{15},\frac{7}{15},1\right\}.
\label{eq:points}
\end{equation}
As illustrated in Figure \ref{fig:quan}(a), with a $b=3$-bit uniform
quantizer, the receiver is unable to distinguish between the transmitted
points $t_1$ and $t_2$, since they both fall in the same quantization
interval. It is expected that a quantizer which can distinguish between
all possible transmitted points would have a better rate region than a
quantizer which fails to do so. Hence, as shown in Fig. \ref{fig:quan}(b),
with $b=3$ bits, a non-uniform quantizer, which distinguishes between all
possible transmitted points, would have a better rate region than what is
achieved by a $b=3$ bit uniform quantizer. In the following subsection, we
propose a non-uniform quantizer for QMAC. We will see that indeed the
proposed non-uniform quantizer enlarges the rate region.

\subsection{A Non-uniform Quantizer for QMAC} \label{sec:nonuni}
In this subsection, we propose a non-uniform quantizer for QMAC.
The function $\mbf{Q}_b(.)$
for the real component of the received signal in the proposed
non-uniform quantizer is

\vspace{-5mm}
{\small
\begin{equation}
 r^I=\mbf{Q}_b(y^I)\Define\left\{\begin{array}{cl}
\hspace{-4cm} +1, & \hspace{-4cm}\zeta(y^I) > (2^{b-1}-1) \vspace{1.5mm} \\
\hspace{-4cm} -1, & \hspace{-4cm} \zeta(y^I) < -(2^{b-1}-1) \vspace{1.5mm} \\
\hspace{-2mm}\frac{\displaystyle1}{\displaystyle2} \left[ \left(
\frac{\displaystyle2\zeta(y^I)}{\displaystyle2^b-1} \right)^p+
\left( \frac{\displaystyle2(\zeta(y^I)+1)}{\displaystyle2^b-1}\right)^p \right]\hspace{-1mm}, & \hspace{-1mm} \mbox{o.w.},
\end{array} \right. \label{eq:Inonuni}
\end{equation}
}

\vspace{-4mm}
where $p\geq1$,
{\small $\zeta(y^I) \Define \left\lfloor \left(\frac{2^b-1}{2} \right) \left( \frac{y^I}{X^I} \right)^{1/p} \right\rfloor$} and $X^I$ defined in (\ref{eq:xiq}).
Likewise, for the imaginary component

\vspace{-4mm}
{\small
\begin{equation}
r^Q=\mbf{Q}_b(y^Q)\Define\left\{\begin{array}{cl}
\hspace{-4cm} +1, & \hspace{-4cm}\zeta(y^Q) > (2^{b-1}-1) \vspace{1.5mm} \\
\hspace{-4cm} -1, & \hspace{-4cm} \zeta(y^Q) < -(2^{b-1}-1) \vspace{1.5mm} \\
\hspace{-3mm}\frac{\displaystyle1}{\displaystyle2} \left[ \left(
\frac{\displaystyle2\zeta(y^Q)}{\displaystyle2^b-1} \right)^p+
\left( \frac{\displaystyle2(\zeta(y^Q)+1)}{\displaystyle2^b-1}\right)^p \right]\hspace{-1mm}, & \hspace{-1mm} \mbox{o.w.}
\end{array} \right.
\label{eq:Qnonuni}
\end{equation}
}

\vspace{-4mm}
where {\small $\zeta(y^Q) \Define \left\lfloor \left(\frac{2^b-1}{2} \right) \left( \frac{y^Q}{X^Q} \right)^{1/p} \right\rfloor$} and $X^Q$ defined in (\ref{eq:xiq}).

Note that the parameter $p$ in (\ref{eq:Inonuni}) and (\ref{eq:Qnonuni})
is a quantizer design parameter, which is used to increase/decrease the
quantization granularity around the origin. It can be seen that the uniform
quantizer in (\ref{eq:unir}), (\ref{eq:unii}) is a special case of this
non-uniform quantizer with $p=1$.

For a {\em fixed} rotation angle $\theta$ and a quantizer resolution of
$b$ bits, the sum rate $R_1(\theta)+R_2(\theta)$ is a function of the
parameter $p$. Since the sum rate is a function of both $p$
and $\theta$, we shall denote it by
\begin{equation}
R_{sum}(\theta,p)\,=\,R_1(\theta,p)+R_2(\theta,p).
\label{rsum}
\end{equation}
For a fixed $\theta$, the
optimal quantizer parameter, $p^{*}(\theta)$, which maximizes the sum rate
is given by
\begin{equation}
p^*(\theta)\,=\,\arg\max_{p:\;p\geq1} R_{sum}(\theta,p).
\label{eq:popt}
\end{equation}
For a fixed $\theta$, we now present a
low-complexity iterative algorithm to find a suboptimum solution to
the maximization problem in (\ref{eq:popt}).

\subsubsection{An Iterative Algorithm to Solve (\ref{eq:popt}) }
Let $p^{(k)}$ and $R^{(k)}=R_{sum}(\theta,p^{(k)})$ denote the value of
$p$ and the sum rate in the $k$th iteration, respectively. The algorithm
starts with $p^{(0)}=1$. In the $(k+1)$th iteration, evaluate
$\tilde{R}^{(k+1)}=R_{sum}(\theta,p^{(k)}+1)$. If
$\tilde{R}^{(k+1)} \geq R^{(k)}$, then go to the next iteration with
$R^{(k+1)}=\tilde{R}^{(k+1)}$ and $p^{(k+1)}=p^{(k)}+1$. If
$\tilde{R}^{(k+1)}<R^{(k)}$, then evaluate (\ref{rsum})
for all values of $p$ in the set

\vspace{-4mm}
{\small
\begin{equation}
P=\left\{p^{(k)}+l\Delta, \; l \in \left\{ \left\lfloor \frac{-1}{\Delta} \right\rfloor,\cdots,-1,0,1,\cdots,\left\lfloor\frac{1}{\Delta}\right\rfloor \right\} \subset \mathbb{Z} \right\},
\label{eq:pset}
\end{equation}

where $\Delta < 1$ is the search granularity of the algorithm. Find
\begin{equation}
\tilde{l} \, = \, \arg\max\limits_{l \in \left\{ \left\lfloor \frac{-1}{\Delta} \right\rfloor,\cdots,0,\cdots,\left\lfloor\frac{1}{\Delta}\right\rfloor \right\}} \; R_{sum}(\theta,p^{(k)}+l\Delta).
\end{equation}
}

\vspace{-3mm}
Output $\tilde{p}(\theta)=p^{(k)}+\tilde{l}\Delta$ as the solution and stop.

In the above, the algorithm iteratively increments $p$ in steps of one
until the sum rate can not be further increased after some iteration $k$.
At this point, a finer granularity search is performed in the neighborhood
of $p^{(k)}$. It is observed numerically that $R_{sum}(\theta,p)$ monotonically increases as a function of $p$ for a fixed $\theta$, and hence, with a sufficiently low value of $\Delta$, the value
of $\tilde{p}(\theta)$ is expected to be close to $p^*(\theta)$.
The rotation angle that maximizes $R_{sum}(\theta,\tilde{p}(\theta))$
is then given by
\begin{equation}
\theta'=\arg\max_{\theta} R_{sum}(\theta,\tilde{p}(\theta)).
\end{equation}

\subsubsection{Results and Discussion}
\begin{figure}[h]
\hspace{-0mm}
\epsfysize=7.8cm
\epsfxsize=9.4cm
\hspace{-4mm}
\epsfbox{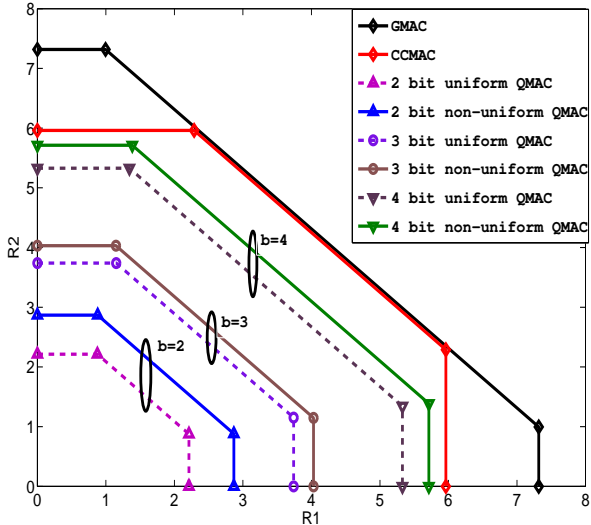}
\vspace{-9.0mm}
\caption{Comparison between the achievable rate regions with uniform and
non-uniform receiver quantization for a two-user MAC. User 1 uses a
64-QAM signal set and User 2 uses the optimally rotated 64-QAM.
SNR per user = 22 dB.}
\label{fig:uq_nuq_64}
\end{figure}

\begin{figure}[h]
\epsfysize=7.8cm
\epsfxsize=9.4cm
\hspace{-4mm}
\epsfbox{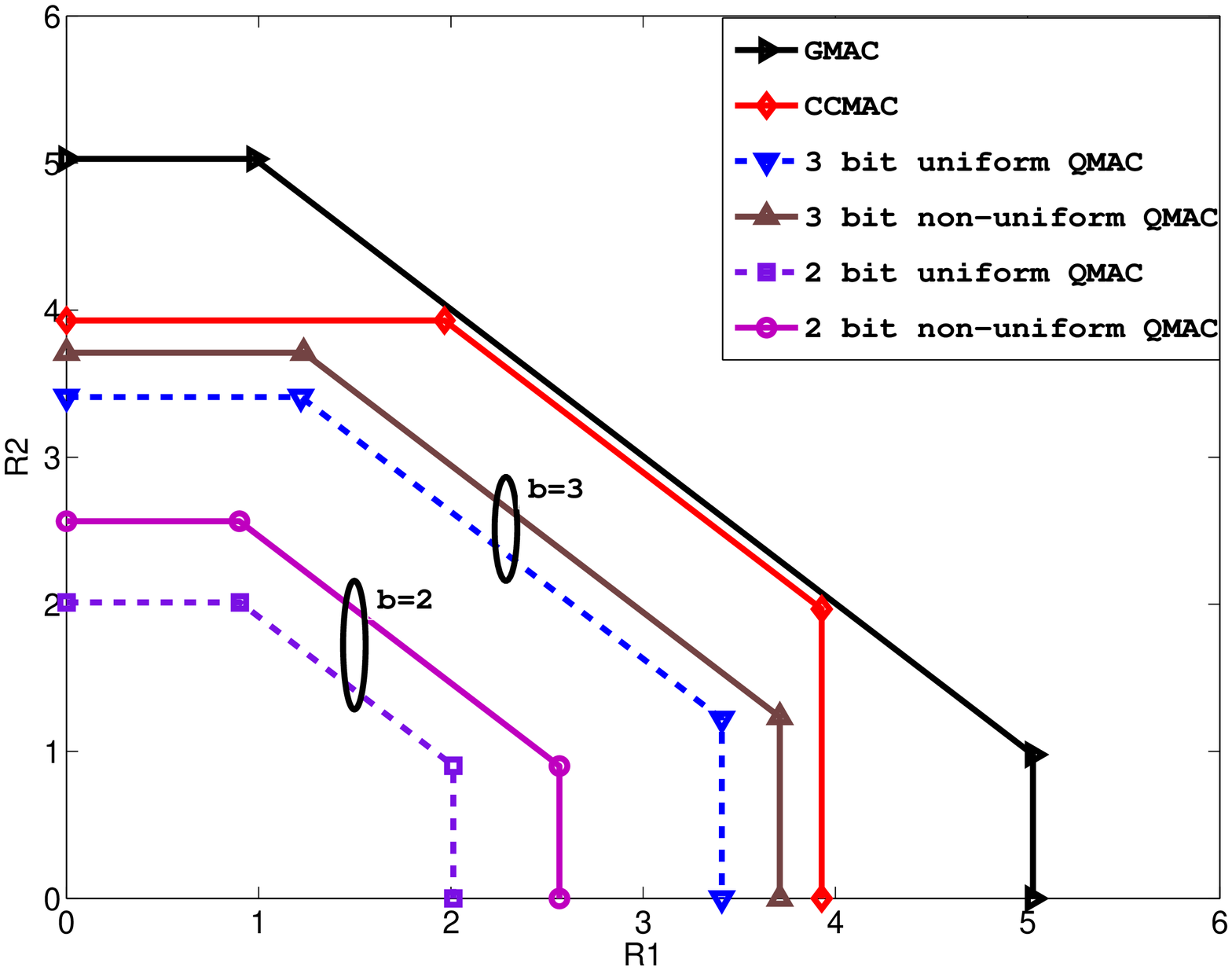}
\vspace{-9.0mm}
\caption{Comparison between the achievable rate regions with uniform and
non-uniform receiver quantization for a two-user MAC. User 1 uses a
16-QAM signal set and User 2 uses the optimally rotated 16-QAM.
SNR per user = 15 dB.}
\label{fig:uq_nuq_16}
\end{figure}

We compute the rate regions achieved by the proposed non-uniform quantizer
for a two-user QMAC with User 1 using a QAM alphabet and User 2 using the
optimally rotated QAM, and compare them with those achieved by the uniform
quantizer. Figure \ref{fig:uq_nuq_64} shows the rate regions for 64-QAM at
SNR per user = 22 dB, and Fig. \ref{fig:uq_nuq_16} shows the rate
regions for 16-QAM at SNR per user = 15 dB. From Fig. \ref{fig:uq_nuq_64},
we observe that the maximum achievable sum rate with a $b=2$-bit uniform
quantizer is 3.0891 bits, whereas a $b=2$-bit non-uniform quantizer
achieves a max. sum rate of 3.7486 bits, which is a $21.35\%$ increase in 
the max. sum rate. This shows that significant enlargement in the rate 
region is achieved with non-uniform quantization compared to uniform 
quantization. Table \ref{tab1} presents a summary of the observations 
from Figs. \ref{fig:uq_nuq_64} and \ref{fig:uq_nuq_16}.

\begin{table}[h]
\begin{center}
\begin{tabular}{|p{0.925cm}|p{13.5mm}|c|p{7.5mm}|m{8.0mm}|}
\hline
 & Uniform & \multicolumn{2}{c|}{Non-uniform} &  \\
\# quant. & Quantizer &  \multicolumn{2}{c|}{Quantizer} & \% gain\\ \cline{2-4}
bits, $b$&$R_{sum}(\theta^{opt})$ & $R_{sum}(\theta',\tilde{p}(\theta'))$ & $\tilde{p}(\theta')$ & \\
&(bits)&(bits)&&\\ \hline \hline
\multicolumn{5}{|p{7.25cm}|}{User 1: 16-QAM, User 2: Optimally rotated 16-QAM 
(Fig. \ref{fig:uq_nuq_16})} \\ \hline
$b=2$ & 2.9144 & 3.4675 & 2.9 & 18.98\\ \hline
$b=3$ & 4.6290 & 5.0787 & 1.3 & 09.71 \\ \hline \hline
\multicolumn{5}{|p{7.25cm}|}{User 1: 64-QAM, User 2: Optimally rotated 64-QAM 
(Fig. \ref{fig:uq_nuq_64})}  \\ \hline
$b=2$ & 3.0891 & 3.7486 & 3.2 & 21.35\\ \hline
$b=3$ & 4.8910 & 5.1790 & 1.6 & 05.89 \\ \hline
\end{tabular}
\caption{Comparison of the maximum achievable sum rates with uniform and non-uniform receiver quantization for a two-user QMAC }
\label{tab1}
\vspace{-8mm}
\end{center}
\end{table}

\section{Conclusions}
\label{sec_concl}
We studied the achievable rate region of quantized broadcast and MAC
channels. We showed that the capacity region expressions known for a 
GBC can not be used as such for QBC as the channel is no more degraded.
We proposed a new achievable rate region for two-user QBC based on two 
different schemes. We studied the proposed achievable rate region of 
two-user QBC when both the users employ uniform receiver quantization. 
We studied the effect of rotating one of the user's input alphabet 
on the proposed achievable rate region of QBC.
Further, we investigated the effect of receiver quantization on the 
achievable rate region of QMAC. Low-precision quantization was shown 
to significantly degrade the rate region. Uniform quantization was 
found to result in significant rate loss due to the dense distribution 
of the sum signal set near the origin. We proposed a non-uniform 
quantizer that achieved significant enlargement of the achievable 
rate region compared to that with a uniform quantizer. 

\bibliographystyle{IEEE}

\begin{thebibliography}{99}
\bibitem{walden99}
R. H. Walden, {\em ADC Survey and Analysis}, {\em IEEE Jl. Sel. Areas in
Commun.}, vol. 17, no. 4, pp. 539-550, April 1999.

\bibitem{ashu}
G. Middleton and A. Sabharwal, ``On the impact of finite receiver resolution
in fading channels,'' {\em Allerton Conf. on Communication, Control and
Computing}, September 2006.

\bibitem{mezg}
A. Mezghani, M. S. Khoufi, and J. A. Nossek, ``Maximum likelihood detection
for quantized MIMO systems,'' {\em The Intl. Workshop on Smart Antennas
(WSA'2008)}, pp. 278-284, Darmstadt, February 2008.

\bibitem{singh09}
J. Singh, O. Dabeer, and U. Madhow, ``On the limits of communication with
low-precision analog-to-digital conversion at the receiver,'' {\em IEEE
Tran. Commun.}, vol. 52, no. 12, pp. 3629-3639, December 2009.

\bibitem{gamal}
Gamal, A.E., ``The capacity of a class of broadcast channels,'' {\em IEEE
Trans. Inform. Theory}, vol. 25, pp. 166-169, March 1979.

\bibitem{naveen}
N. Deshpande and B. Sundar Rajan, ``Constellation constrained capacity
of two-user broadcast channels,'' {\em Proc. IEEE GLOBECOM'09}, Honolulu,
November-December 2009.

\bibitem{biglieri}
E. Biglieri, {\em Coding for wireless channels}, Springer-Verlag,
NY, 2005.

\bibitem{harsh08}
J. Harshan and B. Sundar Rajan, ``Finite signal-set capacity of two-user
Gaussian multiple access channel,'' {\em Proc. IEEE ISIT'2008},
pp. 1203-1207, July 2008.

\bibitem{cover99}
T. M. Cover and J. A. Thomas, {\em Elements of Information Theory},
2nd Edition, Wiley Series in Telecommmun. and Sig. Proc., 1999.

\bibitem{cover_bc}
T. M. Cover, ``Broadcast channels,'' {\em IEEE Trans. Inform. Theory},
vol. IT-18, no. 1, pp. 2-14, January 1972.

\bibitem{boyd}
S. Boyd and L. Vandenberghe, {\em Convex Optimization},
{\em Cambridge University Press}, 2004.

\bibitem{qmac}
Suresh Chandrasekaran, Saif K. Mohammed, and A. Chockalingam, ``On the
capacity of quantized Gaussian MAC channels with finite input alphabet,''
{\em to appear in Proc. IEEE ICC'2011}, Kyoto, June 2011.

\end{thebibliography}

\end{document}